# New approach to model the yield strength of body centered cubic solid solution refractory high entropy alloys

Ali Shafiei

Metallurgy Group, Niroo Research Institute (NRI), Tehran 14665-517, Iran

E-mail: alshafiei@nri.ac.ir

**Abstract**

A data fitting approach is used in the present work for modeling the yield strength of single-phase body centered cubic (bcc) refractory high entropy alloys (RHEAs) in the Al-Hf-Nb-Mo-Ta-Ti-V-Zr system. It is suggested that a polynomial equation could be used for modeling the strength of solid solution alloys where an experimental dataset containing more than 80 data is used for fitting and obtaining optimized polynomial coefficients. The comparison results show that the predictions are in good agreement with experimental data indicating that the proposed polynomial could be used for modeling the strength of RHEAs. To show how the developed polynomial can be applied in designing of RHEAs, the alloy system Hf-Mo-Nb-Ta-Ti-V-Zr is selected as an example and the strength of designed RHEAs within this system is investigated by the developed polynomial. The results show that the strength of alloys within this system increases by increasing the amount of Mo and Zr while the strength of alloys decreases by increasing the amount of Ti. Moreover, the results indicate that the strength of alloys increases with increasing the values of parameters atomic size difference (ASD) and valence electron concentration (VEC). The developed polynomial can be considered as a straightforward method for assessing the strength of solid solution RHEAs, although it does not explain the mechanisms involved in the strengthening of alloys.

## 1. Introduction

High entropy alloys (HEAs) or multi-principal-element alloys are a new group of metallic alloys with encouraging functional and mechanical properties [1-4]. The main difference between HEAs and traditional alloys is that HEAs contain multiple principal elements while traditional alloys are usually based on one dominant element [1-4]. HEAs which are based on refractory elements such as Hf, Nb, Mo, Ta, W, V and Zr are called refractory high entropy alloys (RHEAs) after Senkov et al. [5-6]. RHEAs have shown very promising high-temperature mechanical properties and are considered as "future materials for high-temperature structural applications beyond Ni-based superalloys” [5]. A detailed review regarding the microstructure and properties of RHEAs can be found in Ref. [5].

The final goal of many research works in the field of HEAs is to identify alloys with desired properties. However, because HEAs contain several principal elements, a great number of alloys need to be examined in this regard. Examinating a great number of alloys is not possible in practice. Therefore, it is very important to develop convenient methods and models which can predict the microstructure and properties of HEAs. In this regard, several research works have focused on developing models for predicting the microstructure or mechanical properties of HEAs [1-5]. Although these approaches have been successfully applied, they have their own limitations. The objective of the present work is to introduce a new methodology (data-based modeling and fitting) for predicting the properties of HEAs. In the present work, the approach is applied for predicting the yield strength of body centered cubic (bcc) solid solution RHEAs in the Al-Hf-Nb-Mo-Ta-Ti-V-Zr system. Although the introduced approach (data-based modeling and fitting) do not explain the mechanisms involved in the strengthening of alloys, it can be considered as a straightforward method for assessing the strength of solid solution RHEAs in the Al-Hf-Nb-Mo-Ta-Ti-V-Zr system. Therefore, the proposed approach can be considered as a valuable tool for designing RHEAs in this alloy system.

## 2. Methodology

### 2. 1. Experimental data

RHEAs (more than 80 alloys) which are shown in Table 1 are used in the preset work. The data in Table 1 are gathered from Refs. [7-37]. All of the alloys in Table 1 were made by vacuum arc melting technique, and a single phase bcc solid solution microstructure was reported for all of them in the as-cast condition. Some Ti rich alloys are also listed in Table 1. These Ti rich alloys are being considered as promising candidates for biomedical applications [35-37]. The compressive yield stress for each alloy in the as-cast condition is shown in the second column. It can be seen that the dataset in Table 1 covers a relatively broad range of chemical compositions including Ti and Nb rich alloys. The composition domain which is covered by alloys in Table 1 is schematically shown in Fig. 1. In addition to the experimental data in Table 1, following estimations [27, 38] are also used for the yield strength of bcc elements and binary systems: $\sigma_{Mo}$ = 450 MPa [27], $\sigma_{Nb}$ = 250 MPa [27], $\sigma_{Ta}$ = 350 [27], $\sigma_{V}$ =300 MPa [27], $\sigma_{NbTa}$ =320 MPa [27, 38], $\sigma_{MoTa}$ = 1200 MPa [27, 39], and $\sigma_{VNb}$ = 800 MPa [40].

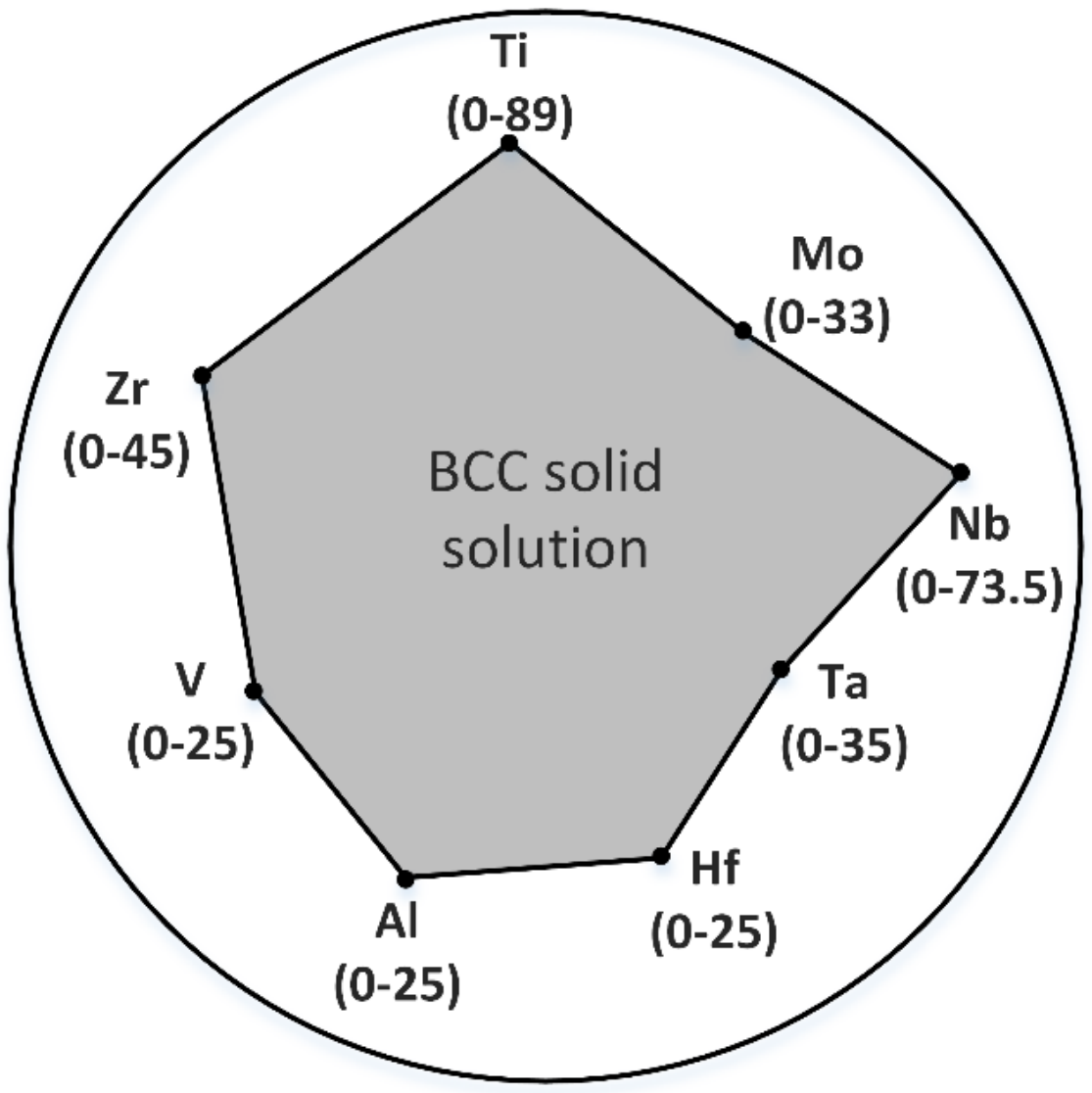

Fig. 1 Composition domain which is covered by alloys in Table 1 containing Ti and Nb rich alloys

Table 1. Chemical compositions of investigated alloys and the compressive yield stress (*S*) for each alloy in the as-cast condition. All of the alloys have a single phase bcc solid solution microstructure in the as-cast condition

| **Alloy** | ***S*(MPa)** | **Al (at %)** | **Hf (at %)** | **Mo (at %)** | **Nb (at %)** | **Ta (at %)** | **Ti (at %)** | **V (at %)** | **Zr (at %)** |
|---|---|---|---|---|---|---|---|---|---|
| HfNbTaTiZr [7] | 1073 | 0 | 20 | 0 | 20 | 20 | 20 | 0 | 20 |
| $Al_{0.3}$HfNbTaTiZr [7] | 1188 | 5.5 | 19 | 0 | 18.9 | 18.9 | 18.9 | 0 | 18.9 |
| $Al_{0.5}$HfNbTaTiZr [7] | 1302 | 9.1 | 18 | 0 | 18.2 | 18.18 | 18.2 | 0 | 18.2 |
| $Al_{0.75}$HfNbTaTiZr [7] | 1415 | 13 | 17 | 0 | 17.4 | 17.4 | 17.4 | 0 | 17.4 |
| HfMo0NbTaTiZr [8] | 1015 | 0 | 20 | 0 | 20 | 20 | 20 | 0 | 20 |
| $HfMo_{0.25}$NbTaTiZr [8] | 1112 | 0 | 19 | 4.8 | 19.1 | 19.05 | 19.1 | 0 | 19.1 |
| $HfMo_{0.5}$NbTaTiZr [8] | 1317 | 0 | 18 | 9.1 | 18.2 | 18.18 | 18.2 | 0 | 18.2 |
| $HfMo_{0.75}$NbTaTiZr [8] | 1373 | 0 | 17 | 13 | 17.4 | 17.39 | 17.4 | 0 | 17.4 |
| HfMoNbTaTiZr [8] | 1512 | 0 | 17 | 17 | 16.7 | 16.67 | 16.7 | 0 | 16.7 |
| HfNbTiZr [9] | 879 | 0 | 25 | 0 | 25 | 0 | 25 | 0 | 25 |
| TiHfZrTaNb [10] | 905 | 0 | 20 | 0 | 20 | 20 | 20 | 0 | 20 |
| $Hf_{0.5}Nb_{0.5}Ta_{0.5}Ti_{1.5}Zr$ [11] | 903 | 0 | 13 | 0 | 12.5 | 12.5 | 37.5 | 0 | 25 |
| AlNbTaTi [12] | 1150 | 25 | 0 | 0 | 25 | 25 | 25 | 0 | 0 |
| HfMoTaTiZr [13] | 1600 | 0 | 20 | 20 | 0 | 20 | 20 | 0 | 20 |
| HfMoNbTaTiZr [13] | 1512 | 0 | 17 | 17 | 16.7 | 16.67 | 16.7 | 0 | 16.7 |
| NbMoTiWHf [14] | 521 | 0 | 0 | 9.2 | 73 | 0 | 17.8 | 0 | 0 |
| NbMoTiWHf [14] | 719 | 0 | 0 | 12 | 70 | 0 | 18.1 | 0 | 0 |
| NbMoTiWHf [14] | 780 | 0 | 0 | 14 | 67.8 | 0 | 18.1 | 0 | 0 |
| NbMoTiWHf [14] | 1100 | 0 | 0 | 33 | 34.7 | 0 | 32.5 | 0 | 0 |
| $Hf_{0.4}Nb_{1.54}Ta_{1.54}Ti_{0.89}Zr_{0.64}$ [15] | 822 | 0 | 8 | 0 | 30.8 | 30.8 | 17.7 | 0 | 12.7 |
| $Hf_{0.5}Mo_{0.5}NbTiZr$ [16] | 1176 | 0 | 13 | 13 | 25 | 0 | 25 | 0 | 25 |
| HfMoNbTiZr [17] | 1719 | 0 | 20 | 20 | 20 | 0 | 20 | 0 | 20 |
| HfNbTaZr [18] | 1315 | 0 | 25 | 0 | 25 | 25 | 0 | 0 | 25 |
| MoNbTiZr [19] | 1560 | 0 | 0 | 25 | 25 | 0 | 25 | 0 | 25 |
| $TiZrNbMoV_{0.25}$ [19] | 1750 | 0 | 0 | 23.53 | 23.53 | 0 | 23.53 | 5.88 | 23.53 |
| HfMoTaTiZr [20] | 1600 | 0 | 20 | 20 | 0 | 20 | 20 | 0 | 20 |
| HfMoNbTiZr [20] | 1351 | 0 | 20 | 20 | 20 | 0 | 20 | 0 | 20 |
| HfMoNbTaZr [20] | 1524 | 0 | 20 | 20 | 20 | 20 | 0 | 0 | 20 |
| HfMoNbTaTi [20] | 1367 | 0 | 20 | 20 | 20 | 20 | 20 | 0 | 0 |
| $HfNb_{0.5}Mo_{0.5}TiZr$ [21] | 1195 | 0 | 25 | 13 | 12.5 | 0 | 25 | 0 | 25 |
| $HfNb_{0.5}Ta_{0.5}ZrTi$ [21] | 738 | 0 | 25 | 0 | 12.5 | 12.5 | 25 | 0 | 25 |
| $Mo_{1.4}Nb_{1.4}Ta_{1.4}Ti_{0.6}Zr_{0.6}$ [22] | 1275 | 0 | 0 | 26 | 25.9 | 25.93 | 11.1 | 0 | 11.1 |
| $Mo_{0.6}Nb_{0.6}Ta_{0.6}Ti_{1.4}Zr_{1.4}$ [22] | 1160 | 0 | 0 | 13 | 13.1 | 13.05 | 30.4 | 0 | 30.4 |
| $Mo_{0.5}NbTaTi_{1.5}Zr$ [22] | 1050 | 0 | 0 | 10 | 20 | 20 | 30 | 0 | 20 |
| $Mo_{0.3}NbTaTi_{1.7}Zr$ [22] | 1025 | 0 | 0 | 6 | 20 | 20 | 34 | 0 | 20 |
| $Ti_{50-x}Al_xV_{20}Nb_{20}Mo_{10}$ [23] | 900 | 10 | 0 | 10 | 20 | 0 | 40 | 20 | 0 |
| $Ti_{50-x}Al_xV_{20}Nb_{20}Mo_{10}$ [23] | 971 | 15 | 0 | 10 | 20 | 0 | 35 | 20 | 0 |
| $Ti_{50-x}Al_xV_{20}Nb_{20}Mo_{10}$ [23] | 1187 | 20 | 0 | 10 | 20 | 0 | 30 | 20 | 0 |
| MoNbTiV [24] | 1200 | 0 | 0 | 25 | 25 | 0 | 25 | 25 | 0 |
| Al0.25MoNbTiV [24] | 1250 | 5.88 | 0 | 23.53 | 23.53 | 0 | 23.53 | 23.53 | 0 |
| Al0.5MoNbTiV [24] | 1625 | 11.12 | 0 | 22.22 | 22.22 | 0 | 22.22 | 22.22 | 0 |
| Al0.75MoNbTiV [24] | 1260 | 15.8 | 0 | 21.05 | 21.05 | 0 | 21.05 | 21.05 | 0 |
| AlMoNbTiV [24] | 1375 | 20 | 0 | 20 | 20 | 0 | 20 | 20 | 0 |
| MoVTaTi [25] | 1221 | 0 | 0 | 25 | 0 | 25 | 25 | 25 | 0 |
| $HfMo_{0.5}NbTiV_{0.5}$ [26] | 1260 | 0 | 25 | 12.5 | 25 | 0 | 25 | 12.5 | 0 |
| $Mo_{0.1}NbTiV_{0.3}Zr$ [26] | 932 | 0 | 0 | 2.94 | 29.41 | 0 | 29.41 | 8.83 | 29.41 |
| $Mo_{0.3}NbTiV_{0.3}Zr$ [26] | 1312 | 0 | 0 | 8.33 | 27.78 | 0 | 27.78 | 8.33 | 27.78 |
| $Mo_{0.3}NbTiVZr$ [26] | 1289 | 0 | 0 | 6.96 | 23.26 | 0 | 23.26 | 23.26 | 23.26 |
| $Mo_{0.5}NbTiV_{0.3}Zr$ [26] | 1301 | 0 | 0 | 13.14 | 26.32 | 0 | 26.32 | 7.9 | 26.32 |
| $Mo_{0.5}NbTiVZr$ [26] | 1473 | 0 | 0 | 11.12 | 22.22 | 0 | 22.22 | 22.22 | 22.22 |
| $Mo_{0.7}NbTiV_{0.3}Zr$ [26] | 1436 | 0 | 0 | 17.5 | 25 | 0 | 25 | 7.5 | 25 |
| $Mo_{0.7}NbTiVZr$ [26] | 1706 | 0 | 0 | 14.88 | 21.28 | 0 | 21.28 | 21.28 | 21.28 |
| $Mo_{1.3}NbTiV_{0.3}Zr$ [26] | 1603 | 0 | 0 | 28.26 | 21.74 | 0 | 21.74 | 6.52 | 21.74 |
| $Mo_{1.3}NbTiVZr$ [26] | 1496 | 0 | 0 | 24.52 | 18.87 | 0 | 18.87 | 18.87 | 18.87 |
| $Mo_{1.5}NbTiV_{0.3}Zr$ [26] | 1576 | 0 | 0 | 31.26 | 20.83 | 0 | 20.83 | 6.25 | 20.83 |
| TiZrNbV [26] | 1104 | 0 | 0 | 0 | 25 | 0 | 25 | 25 | 25 |
| $TiZrNbV_{0.3}$ [26] | 866 | 0 | 0 | 0 | 30.3 | 0 | 30.3 | 9.1 | 30.3 |
| TiZrNbVMo [26] | 1779 | 0 | 0 | 20 | 20 | 0 | 20 | 20 | 20 |
| $MoNbTiV_{0.3}Zr$ [26] | 1455 | 0 | 0 | 23.25 | 23.25 | 0 | 23.25 | 7 | 23.25 |
| MoNbTaTiV [27] | 1400 | 0 | 0 | 20 | 20 | 20 | 20 | 20 | 0 |
| MoNbTaV [28] | 1500 | 0 | 0 | 25 | 25 | 25 | 0 | 25 | 0 |
| NbTaTiV [29] | 965 | 0 | 0 | 0 | 25 | 25 | 25 | 25 | 0 |
| $Ti_2ZrHf_{0.5}VNb_{0.25}$ [30] | 1115 | 0 | 10.53 | 0 | 5.26 | 0 | 42.11 | 21.05 | 21.05 |
| $Ti_2ZrHf_{0.5}VNb_{0.5}$ [30] | 1065 | 0 | 10 | 0 | 10 | 0 | 40 | 20 | 20 |
| $Ti_2ZrHf_{0.5}VNb_{0.75}$ [30] | 1025 | 0 | 9.5 | 0 | 14.3 | 0 | 38.2 | 19 | 19 |
| $Ti_2ZrHf_{0.5}VNb$ [30] | 980 | 0 | 16.67 | 0 | 16.67 | 0 | 33.34 | 16.67 | 16.65 |
| TaZrNbTi [31] | 1100 | 0 | 0 | 0 | 25 | 25 | 25 | 0 | 25 |
| TiZrNbHf [32] | 1000 | 0 | 25 | 0 | 25 | 0 | 25 | 0 | 25 |
| NbMoTa [33] | 999 | 0 | 0 | 33 | 34 | 33 | 0 | 0 | 0 |
| TiZrNbTa [34] | 925 | 0 | 0 | 0 | 25 | 25 | 25 | 0 | 25 |
| $Ti_{31.67}Zr_{31.67}Nb_{31.66}Ta_5$ [34] | 1075 | 0 | 0 | 0 | 31.66 | 5 | 31.67 | 0 | 31.67 |
| $Ti_{35}Zr_{35}Nb_{25}Ta_5$ [34] | 1000 | 0 | 0 | 0 | 25 | 5 | 35 | 0 | 35 |
| $Ti_{45}Zr_{45}Nb_5Ta_5$ [34] | 850 | 0 | 0 | 0 | 5 | 5 | 45 | 0 | 45 |
| $Ti_{21.67}Zr_{21.67}Nb_{21.66}Ta_{35}$ [34] | 1250 | 0 | 0 | 0 | 21.66 | 35 | 21.67 | 0 | 21.67 |
| $Ti_{15}Zr_{15}Nb_{35}Ta_{35}$ [34] | 1050 | 0 | 0 | 0 | 35 | 35 | 15 | 0 | 15 |
| Ti-Nb-Zr [35] | 720 | 0 | 0 | 0 | 25 | 0 | 57 | 0 | 18 |
| Ti-Nb-Zr [35] | 750 | 0 | 0 | 0 | 22 | 0 | 54 | 0 | 24 |
| Ti-Nb-Zr [35] | 740 | 0 | 0 | 0 | 21 | 0 | 52 | 0 | 27 |
| Ti-Nb-Zr [35] | 715 | 0 | 0 | 0 | 19 | 0 | 50 | 0 | 31 |
| Ti-Mo-Nb [36] | 860 | 0 | 0 | 8 | 3 | 0 | 89 | 0 | 0 |
| Ti-Mo-Nb [36] | 762 | 0 | 0 | 8 | 6 | 0 | 86 | 0 | 0 |
| Ti-Mo-Nb [36] | 708 | 0 | 0 | 9 | 9 | 0 | 82 | 0 | 0 |
| Ti-Mo-Nb [37] | 535 | 0 | 0 | 2 | 26 | 0 | 72 | 0 | 0 |
| Ti-Mo-Nb [37] | 621 | 0 | 0 | 4 | 26 | 0 | 70 | 0 | 0 |
| Ti-Mo-Nb [37] | 634 | 0 | 0 | 6 | 26 | 0 | 68 | 0 | 0 |
| Ti-Mo-Nb [37] | 663 | 0 | 0 | 8 | 26 | 0 | 66 | 0 | 0 |

## 2. 2. Proposed methodology

First the strength of binary solid solution refractory alloys can be considered. Figure 2 shows the critical resolved shear stress ($\tau_{CRSS}$) of Nb-Ta alloy single crystals at room temperature (RT) [38]. It can be seen that the strength of Nb-Ta bcc solid solution alloys can be modeled by a polynomial where the polynomial equation (Eq. 1) can be written as following,

$$\tau_{\mathrm{CRSS}} = 0.8787 \times C_{\mathrm{Nb}} - 0.0062 \times C_{\mathrm{Nb}}^2 + 0.0042 \times C_{\mathrm{Ta}}^2 \,, \qquad (1)$$

where $C_{\mathrm{Nb}}$ and $C_{\mathrm{Ta}}$ show the atomic concentrations of niobium and tantalum in the atomic percent. Therefore, the solid solution strengthening can be modeled by a polynomial equation. As other examples, solid solution strengthening in Mo-Ta [39], Mo-W [39] and V-Nb [40] alloys can be considered. Figure 3 shows the hardness values of Mo-Ta and Mo-W solid solution bcc alloys in homogenized states [39]. It can be observed that the strength of solid solution alloys can be modeled by polynomials as following,

$$H_{\mathrm{Mo-Ta}} = 15.0133 \times C_{\mathrm{Mo}} - 0.1335 \times {C_{\mathrm{Mo}}}^2 + 0.0108 \times {C_{\mathrm{Ta}}}^2, \tag{2}$$

$$H_{\mathrm{Mo-W}} = 4.795 \times C_{\mathrm{W}} + 0.017 \times {C_{\mathrm{Mo}}}^2 - 0.0082 \times {C_{\mathrm{W}}}^2, \tag{3}$$

where $H_{\mathrm{Mo-Ta}}$ is the hardness of Mo-Ta alloys, and $C_{\mathrm{Mo}}$, $C_{\mathrm{Ta}}$ and $C_{\mathrm{W}}$ show the atomic concentration of molybdenum, tantalum and tungsten, respectively.

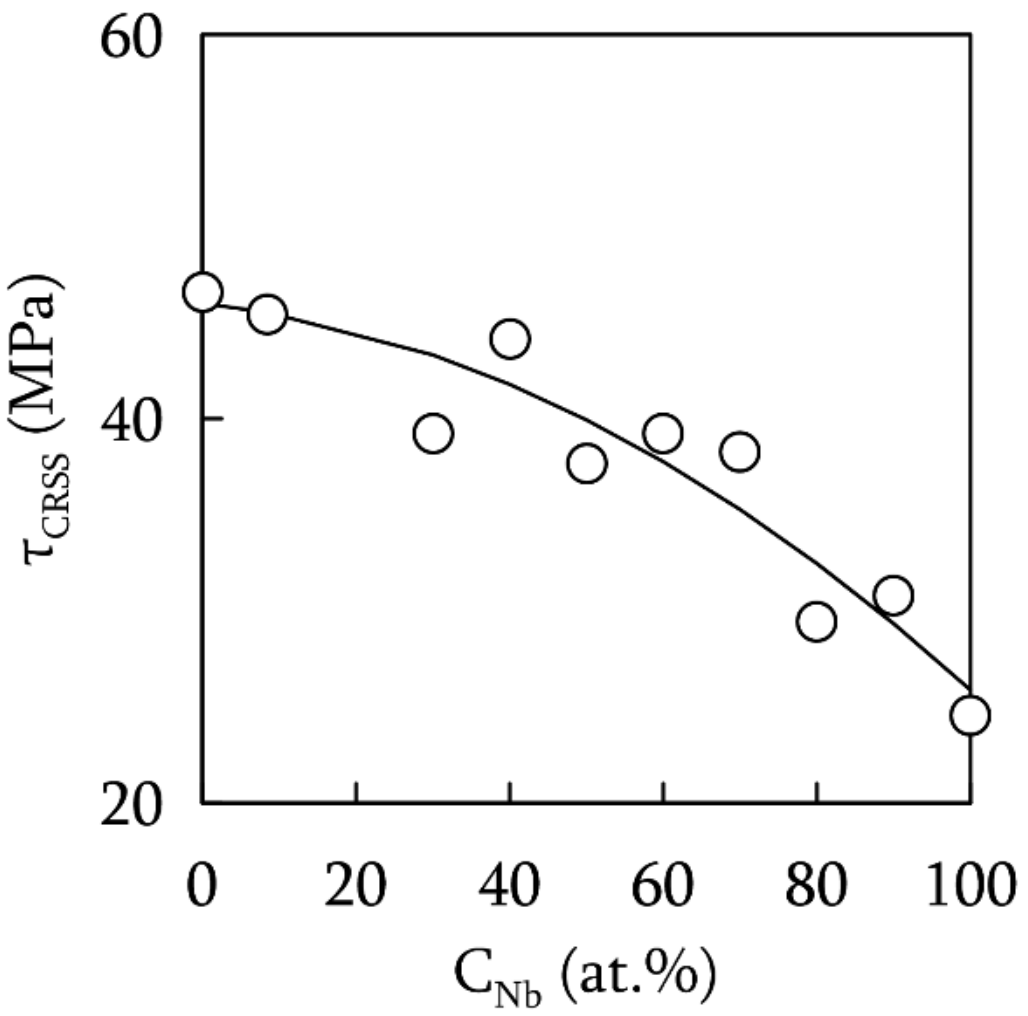

Fig. 2 Experimental data for $\tau_{CRSS}$ of Nb-Ta alloy single crystals at room temperature [38]. The proposed polynomial for modeling the strength can be seen in Eq. (1)

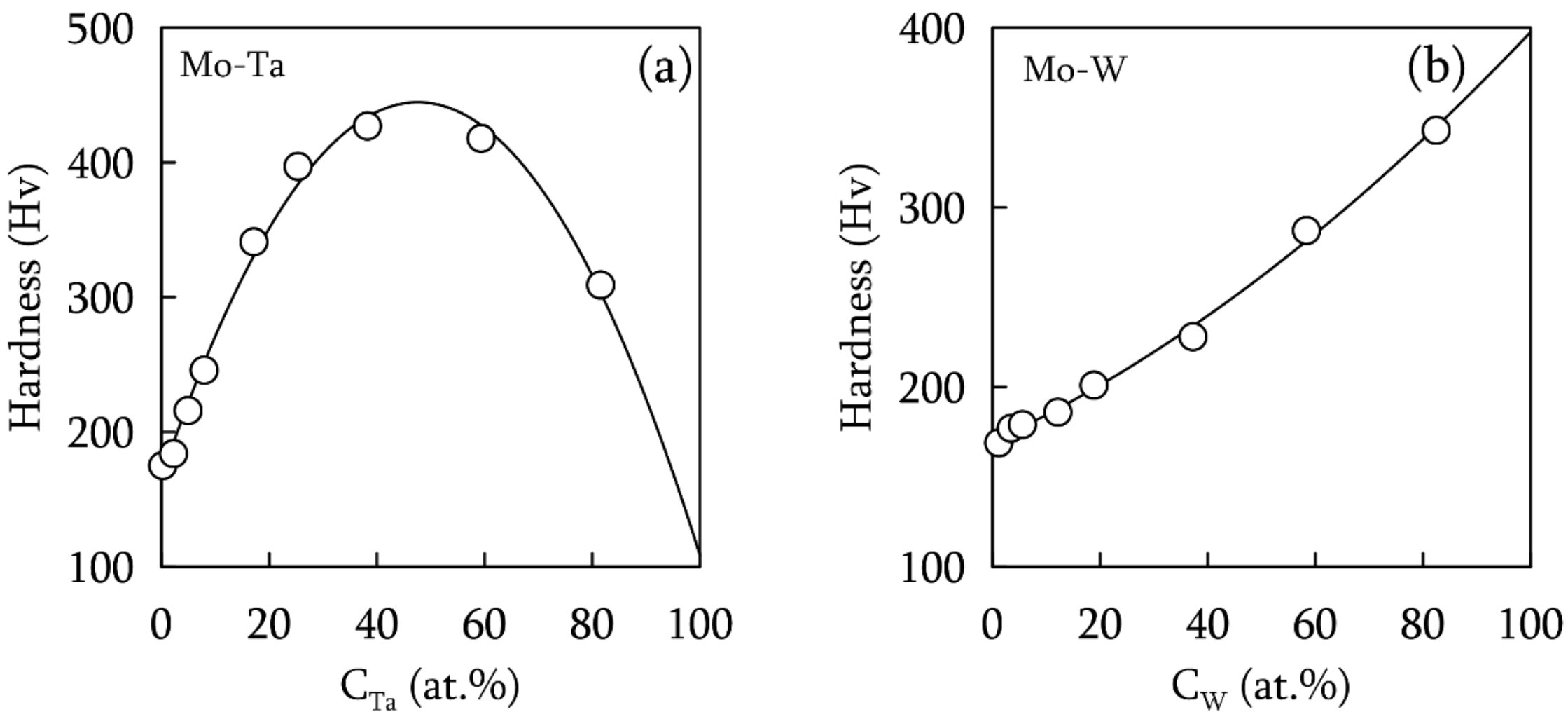

Fig. 3. Experimental data for the hardness (*H*) of **a** Mo-Ta and **b** Mo-W solid solution bcc alloys at room temperature [39]. The proposed polynomials for modeling the hardness can be seen in Eqs. (2) and (3)

According to the results in Figs. 2 and 3, it may be assumed that the strength of solid solution bcc RHEAs can also be modeled by a polynomial. Therefore, the following polynomial (Eq. 4) is proposed in the present work for modeling the compressive yield stress of alloys listed in Table 1

$$S = \sum_{i=1}^{8} (a_i C_i + b_i C_i^2), \qquad (4)$$

where $a_i$ and $b_i$ are coefficients to be determined and $C_i$ is the atomic concentration (at. %) of element *i*. For obtaining $a_i$ and $b_i$, the above equation is fitted to the experimental dataset in Table 1, and the obtained values for coefficients are shown in Table 2.

Table 2. Values of coefficients obtained after fitting the proposed polynomial to the experimental dataset in Table 1

| coefficient | Al | Hf | Mo | Nb | Ta | Ti | V | Zr |
|---|---|---|---|---|---|---|---|---|
| $a_i$ | 15.828 | 17.104 | 36.464 | 7.242 | 10.725 | -0.517 | 19.312 | 34.581 |
| $b_i$ | 0.317 | -0.286 | -0.324 | -0.05 | -0.079 | 0.062 | -0.163 | -0.474 |

The comparison between the predictions and experimental results is shown in Fig. 4. Considering the errors and deviations which exist for the yield stress values in Table 1, it can be seen that a relatively good agreement exists between predictions and experimental results. Therefore, one can conclude that the proposed polynomial can be used for modeling the compressive yield stress of bcc solid solution alloys in the Al-Hf-Nb-Mo-Ta-Ti-V-Zr system. An important point regarding the developed polynomial is that only element concentrations as inputs are needed and no mechanical or structural constants are needed; therefore, it can be applied easily for predicting the strength of RHEAs.

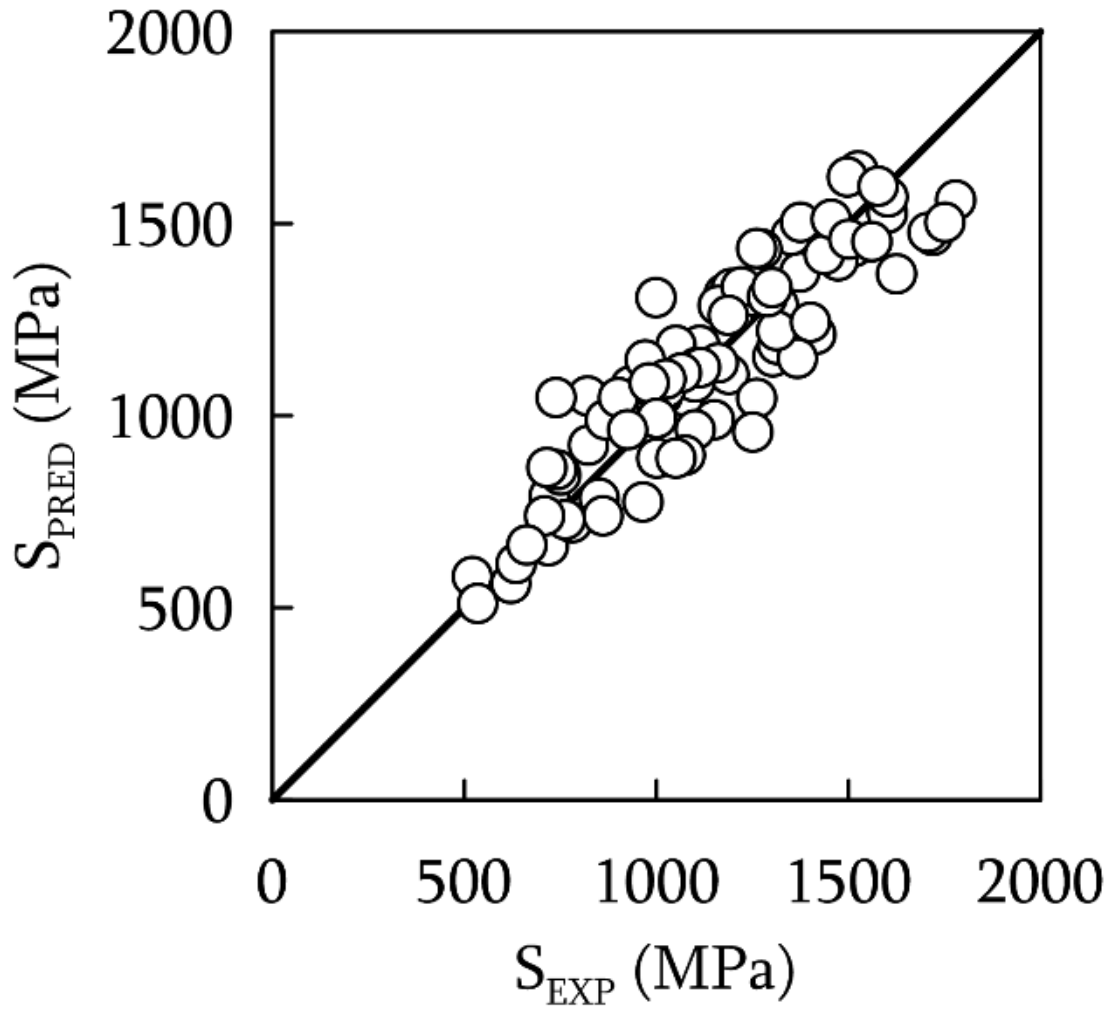

Fig. 4 Comparison between the predictions ($S_{PRED}$) and experimental ($S_{EXP}$) results for the compressive yield stress of alloys in Table 1

## 3. Results and discussions

For verifying the extrapolating ability of developed polynomials, the polynomial is used for predicting the strength of RHEAs in Table 3. A single phase bcc solid solution microstructure in the as-cast condition is reported for all of alloys in Table 3 [35-39]. Because strength values in Table 3 were obtained via tensile or microhardness tests, inconsistencies may be expected between the predictions and experimental results. Furthermore, the Hf and V content of ternary alloys in Table 3 are out of the composition domain in Fig. 1. Therefore, the developed polynomial cannot be used for these alloys, and the predictions results may not be accurate. Nevertheless, the polynomial is applied for these alloys as well.

Table 3. Experimental data (as-cast condition) used for evaluating the extrapolating ability of developed polynomials

| Alloy | as-cast structure | tensile yield stress (MPa) | hardness (HV) | $S_{PRED.}$ (MPa) |
|---|---|---|---|---|
| Nb [41] | bcc | 188 | - | 224.2 |
| NbTi [41] | bcc | 354 | - | 367 |
| NbTiZr [41, 42] | bcc | 749 [41]-956 [42] | 295 [42] | 864.2 |
| NbTiHf [41] | bcc | 613 | - | 492.2 |
| NbTiZrHf [41] | bcc | 783 | - | 992.9 |
| NbTaTiZr [41] | bcc | 876 | - | 962.8 |
| NbTaTiHf [41] | bcc | 762 | - | 643.4 |
| NbTaZrHf [41] | bcc | 1046 | - | 1186 |
| NbTaTiZrHf [41] | bcc | 1142 | - | 1052 |
| HfNbTaTiZr [42] | bcc | 1155 | 359 | 1052 |
| NbTaTiZr [42] | bcc | 1144 | 358 | 962.8 |
| $Nb_{1.5}TaTiZr_{0.5}$ [42] | bcc | 822 | 294 | 804.2 |
| NbTa [43] | bcc | 246 | - | 575.9 |
| HfNbTa [43] | bcc | 847 | - | 716.2 |
| TiNbTa [42-43] | bcc | 620 [42]- 478 [43] | 246 [42] | 505.2 |
| TiHfNbTa [43] | bcc | 663 | - | 643.4 |
| AlNbTaTi [44] | bcc | - | 458[1] | 988.5 |
| MoNbTaTi [44] | bcc | - | 431[2] | 1104 |
| HfNbTaTi [44] | bcc | - | 270[2] | 643.4 |
| AlMoNbTi [44] | bcc | - | 509[2] | 1479 |
| $Ti_{38}V_{15}Nb_{23}Hf_{24}$ [45] | bcc | 800 | - | 709.2 |
| $NbTiVZr_{0.5}$ [46] | bcc | - | 340[2] | 1018 |
| NbTiVZr [46] | bcc | - | 355[2] | 1125 |
| $NbTiVZr_2$ [46] | bcc | - | 350[2] | 1085 |
| $Ti_{45}Zr_{25}Nb_{25}Ta_5$ [47] | bcc | 790 | - | 872.6 |
| $Ti_{40}Zr_{25}Nb_{25}Ta_{10}$ [47] | bcc | 910 | - | 896.4 |
| $Ti_{35}Zr_{25}Nb_{25}Ta_{15}$ [47] | bcc | 1075 | - | 919.4 |
| $Ti_{30}Zr_{25}Nb_{25}Ta_{20}$ [47] | bcc | 1150 | - | 941.5 |
| TaZrHfTi [48] | bcc | 1356 | - | 1062 |
| HfNbTiZr [49] | bcc | 636 | - | 992.9 |
| TiVTa [50] | bcc | 801 | - | 782.1 |
| $Ti_{40}V_{33}Ta_{27}$ [50] | bcc | 926 | - | 770.8 |
| $Ti_{45}V_{20}Ta_{35}$ [50] | bcc | 838 | - | 702.5 |
| $Ti_{48}Zr_{20}Hf_{15}Al_{10}Nb_7$ [51] | bcc | 675 | - | 1051 |
| $Ti_{47}Zr_{20}Hf_{15}Al_{10}Nb_8$ [51] | bcc | 745 | - | 1052 |
| $Ti_{46}Zr_{20}Hf_{15}Al_{10}Nb_9$ [51] | bcc | 700 | - | 1053 |
| $Zr_{50}Ti_{35}Nb_{15}$ [52] | bcc | 657 | - | 699.7 |
| $Ta_{0.7}HfZrTi$ [34] | bcc | 1046 | - | 1048 |
| $Ta_{0.8}HfZrTi$ [34] | bcc | 1120 | - | 1054 |
| TaHfZrTi [34] | bcc | 1367 | - | 1062 |
| $Nb_{0.4}HfZrTi$ [34] | bcc | 744 | - | 979.7 |
| $Nb_{0.8}HfZrTi$ [34] | bcc | 718 | - | 993.7 |
| NbHfZrTi [34] | bcc | 728 | - | 992.9 |
| $Ta_{0.4}Nb_{0.4}HfZrTi$ [34] | bcc | 800 | - | 1038 |
| $Ta_{0.5}Nb_{0.5}HfZrTi$ [34] | bcc | 850 | - | 1048 |
| $Ta_{0.6}Nb_{0.6}HfZrTi$ [34] | bcc | 880 | - | 1053 |
| $Ta_{0.8}Nb_{0.8}HfZrTi$ [34] | bcc | 890 | - | 1056 |
| TaNbHfZrTi [34] | bcc | 1200 | - | 1052 |

1-The compressive yield stress for the AlNbTaTi alloy in as-cast condition is reported to be 1150 MPa [12].

2- For converting microhardness values to yield stress, the relation $S=3H$ can be used.

According to the results in Table 3, it can be seen that a relatively good agreement exists for quaternary and quinary alloy systems, but strength predictions for binary and ternary alloy systems are not accurate. That is probably because few experimental data related to these alloy systems are used for developing the polynomial. Furthermore, the V and Hf content of ternary alloys (33 at.%) is out of the composition domain in Fig. 1. So, the predictions are not accurate for these alloys. If more experimental data for binary and ternary alloy systems can be provided, then a more accurate polynomial and, as a result, more accurate predictions can be obtained. According to the obtained results, it can be concluded that the developed polynomial can predict the strength of bcc RHEAs (not binary and ternary alloy systems) with a reasonable accuracy and therefore it can be considered as an easy-to-use tool for designing RHEAs. It should be noted that although the approach used here is basically a curve fitting exercise and it does not consider the mechanisms involved in the strengthening of alloys, it can be considered as a straightforward method for assessing the strength of solid solution RHEAs in the Al-Hf-Nb-Mo-Ta-Ti-V-Zr system.

## 4. Alloy design

To show how the developed polynomial can be applied in designing RHEAs, the alloy system Hf-Mo-Nb-Ta-Ti-V-Zr is selected as an example, and the strength of RHEAs within this system is predicted by a developed polynomial. In this regard, an iterative MATLAB code is written and more than 40000 alloys with chemical compositions in the following range are examined: 5 at % < $C_{Hf}$ < 20 at %, 5 at % < $C_{Mo}$ < 20 at %, 5 at % < $C_{Nb}$ < 20 at %, 5 at % < $C_{Ta}$ < 20 at %, 5 at % < $C_{Ti}$ < 20 at %, 5 at % < $C_V$ < 20 at %, and 5 at % < $C_{Zr}$ < 20 at %. The relations between the concentrations of constituent elements and the strength of designed alloys in the Hf-Mo-Nb-Ta-Ti-V-Zr system are shown in Fig. 5. It can be observed that the strength increases with increasing the Mo and Zr contents and decreases with increasing the Ti content.

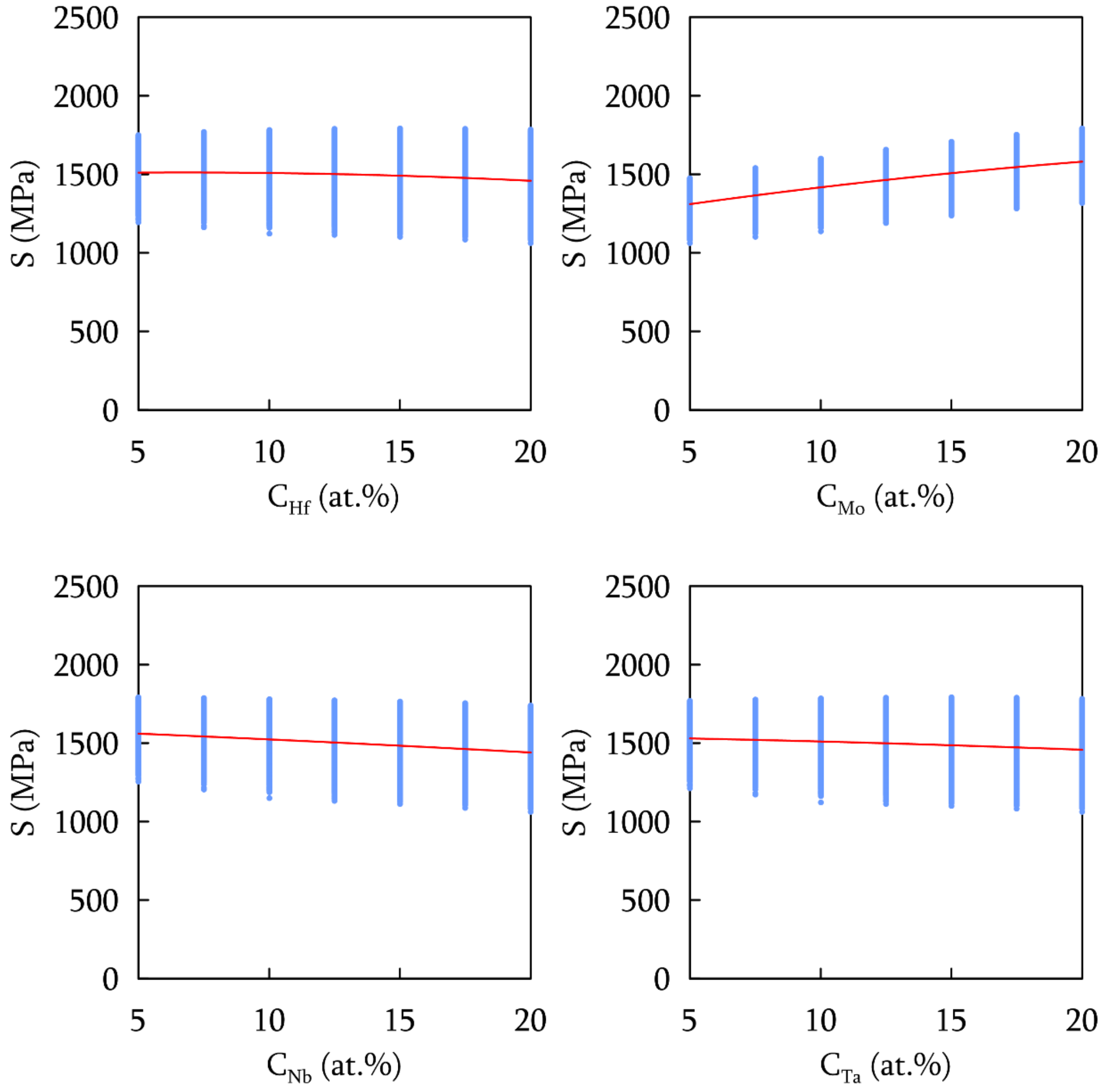

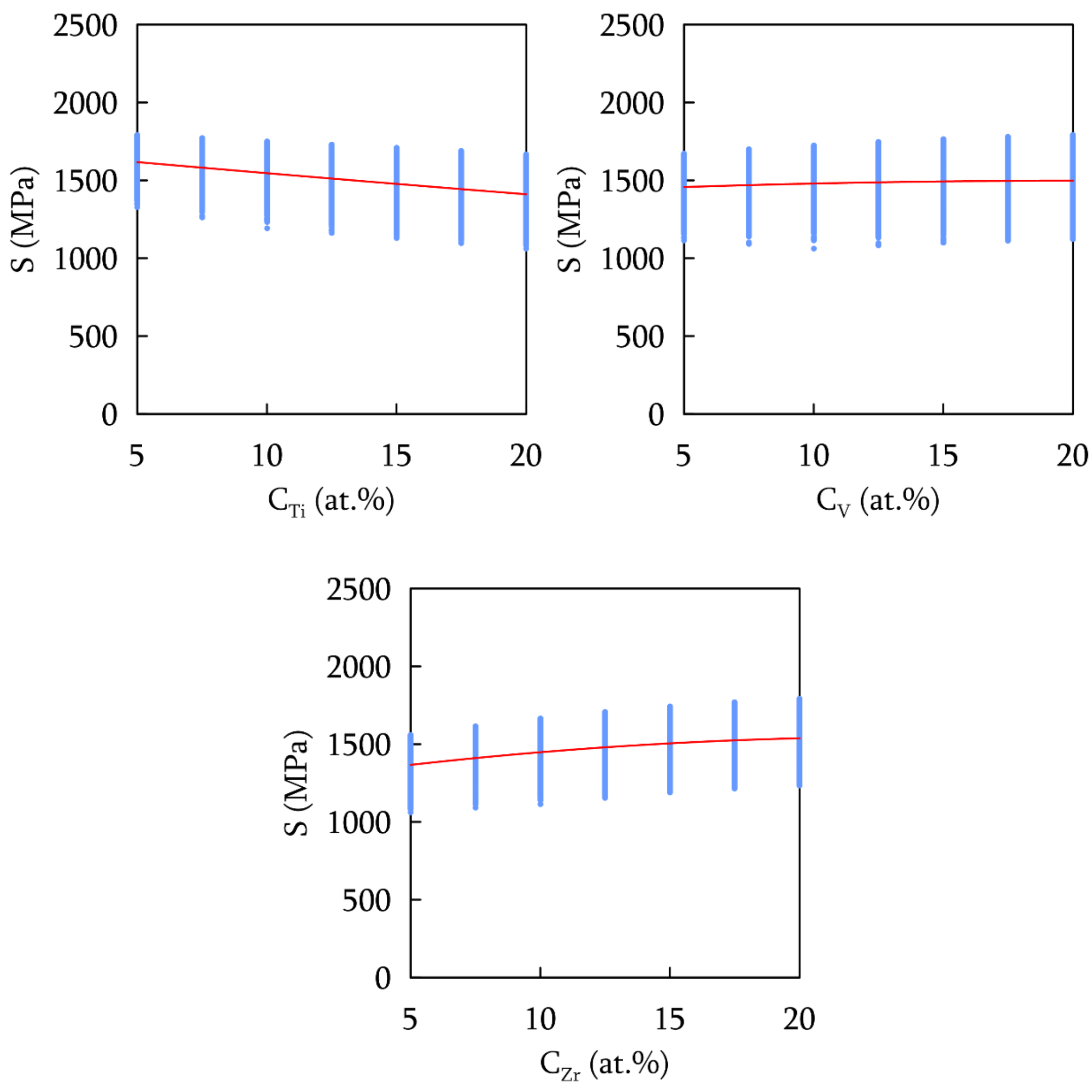

Fig. 5 Effects of constituent elements on the strength of designed RHEAs in the Hf-Mo-Nb-Ta-Ti-V-Zr system (the trends are shown with red lines)

In general, two approaches can be considered for explaining the effect of alloying elements in HEAs [5]. The first approach is according to the Labusch theory [5, 53] which models the solid solution strengthening mechanism by considering the interactions between the dislocations and the non-homogeneities of a lattice. This approach is used in Refs. [8, 12, 20, 29, 54-56]. In the second approach, the non-homogeneities of the lattice are considered within the core of dislocations and the solid solution strengthening mechanism is explained by enhanced Peierls-Nabarro stress ($\sigma_{\mathrm{PN}}$) or friction stress ($\sigma_{\mathrm{f}}$) of the lattice. This approach is used in Refs. [41, 57-60].

The effect of Mo and Zr on the strength of alloys may be explained by considering the Labusch theory [5, 53]. According to the Labusch theory, the strengthening degree of an element depends on the atomic size and elastic modulus mismatches between the solute and solvent, and the strengthening increases with increasing the amount of mismatches. Mo has a relatively low atomic radius in comparison with other constituent elements in the Hf-Mo-Nb-Ta-Ti-V-Zr system ($r_{Mo}$= 0.1362 nm, $r_{avg}$= 0.1454 nm) (Table 4). Furthermore, Mo has the highest Young's modulus among the constituent elements (Table 4). Therefore, when Mo is introduced to the lattice, high atomic size and elastic modulus mismatches are expected which can cause a high degree of strengthening effect for Mo. This may be the reason for the strengthening effect observed for Mo in Fig. 5. On the other hand, Zr has the highest atomic radius among the constituent elements which can cause high atomic size mismatches leading to the strengthening effect for Zr as it can be seen in Fig. 5.

Table 4. Elements properties

| Element | Crystal structure (RT) [3] | Young's Modulus (GPa) [3] | Shear Modulus (GPa) [20] | Yield stress (Polycrystal) (MPa) | VEC [3] | Atom radius (pm) [3] |
|---|---|---|---|---|---|---|
| Hf | hcp | 78 | 30 | 200 [48] | 4 | 157.75 |
| Mo | bcc | 329 | 120 | 438 [27] | 6 | 136.26 |
| Nb | bcc | 105 | 38 | 240 [27] | 5 | 142.9 |
| Ta | bcc | 186 | 69 | 345 [27] | 5 | 143 |
| Ti | hcp | 116 | 44 | 195 [27] | 4 | 146.15 |
| V | bcc | 128 | - | 310 [27] | 5 | 131.6 |
| Zr | hcp | 68 | 33 | 160 [48] | 4 | 160.25 |

The softening behavior of Ti cannot be explained by considering the atomic radius or elastic modules of Ti. One speculation for the softening effect observed for Ti could be its effect in lowering the Peierls-Nabarro stress (PN stress) as a result of reducing the directionality of the electron bondings due to the low valence electron concentration (VEC) of Ti [39, 61-63]. As it can be seen in Table 4, hexagonal close packed (hcp) metals with VEC of 4 have in general lower yield stresses than that of bcc metals with VEC of 5. Furthermore, it can be seen that Mo with VEC of 6 has the highest yield stress in comparison to other elements. So, it may be concluded that a relationship exists between VEC and the strength of alloys. This hypothesis is examined for designed RHEAs, and Fig. 6 shows the relation between VEC and compressive yield stress of designed alloys. It can be seen that a trend exists between VEC and the compressive yield stress of alloys and strength increases with increasing VEC. Therefore, the softening behavior of Ti might be due to its low VEC and its effect on lowering the PN stress. Softening behavior of Ti is reported for Al-Cr-Nb-Ti alloys as well [64-66]. It is reported that Ti increases the fracture toughness of Al-Cr-Nb-Ti alloys by increasing dislocation mobility through a reduction of the PN energy and stress [64-66].

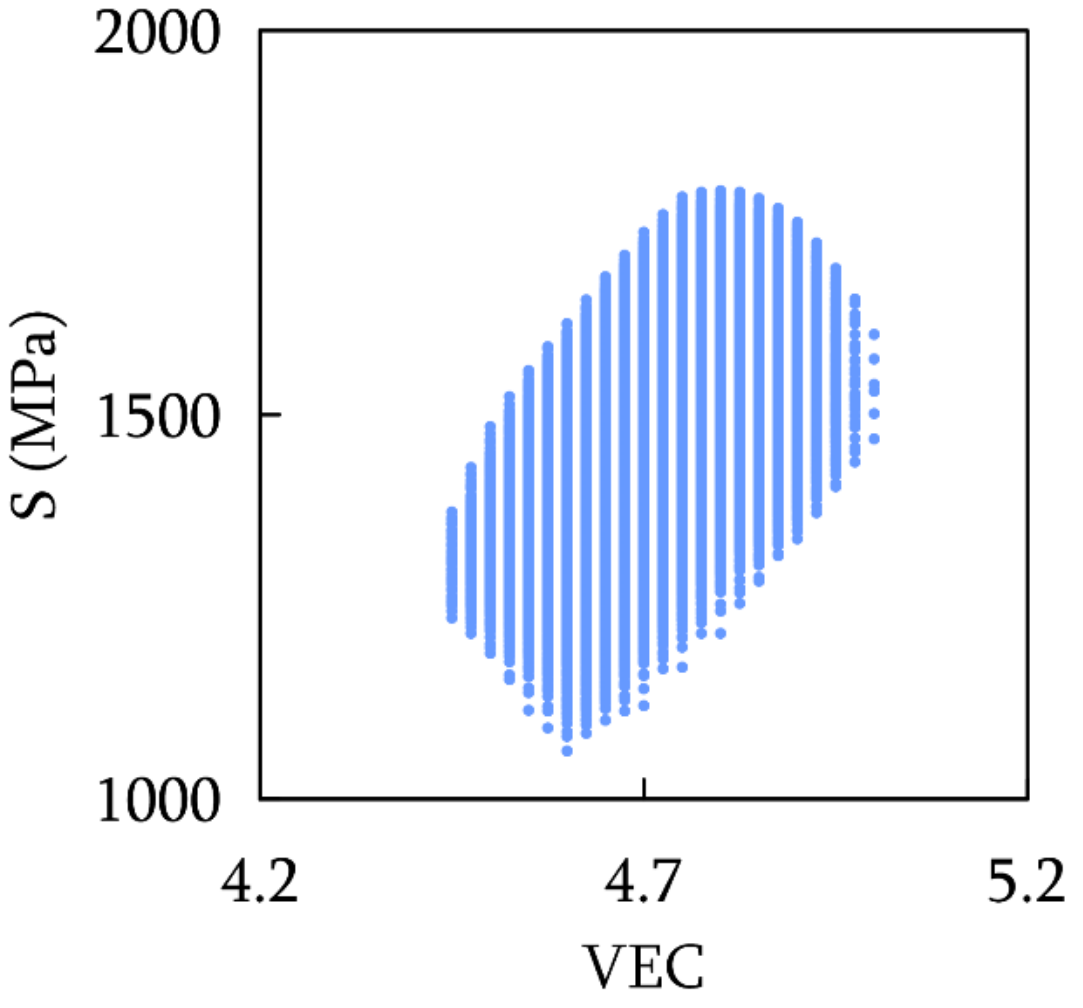

Fig. 6 Relation between VEC and predicted compressive yield stress of designed alloys

The solid solution strengthening effect in HEAs can also be assessed by atomic size difference (ASD) [5] which can be evaluated by Eq. (5) [5]

$$ASD = \sqrt{\sum_{i=1}^{N} C_i \left(1 - \frac{r_i}{\bar{r}}\right)^2}, \qquad (5)$$

where $C_i$ and $r_i$ are the atomic fraction and atomic radius of element *i* respectively, and $\bar{r}$ is the composition-weighted average atomic radius ($\bar{r} = \sum_{i=1}^{N} C_i r_i$). The relation between ASD and the compressive yield stress of designed alloys is shown in Fig. 7. A correlation exists between ASD and the compressive yield stress of alloys, and the strength increases with increasing ASD. Therefore, ASD should be maximized for designing strong solid solution alloys.

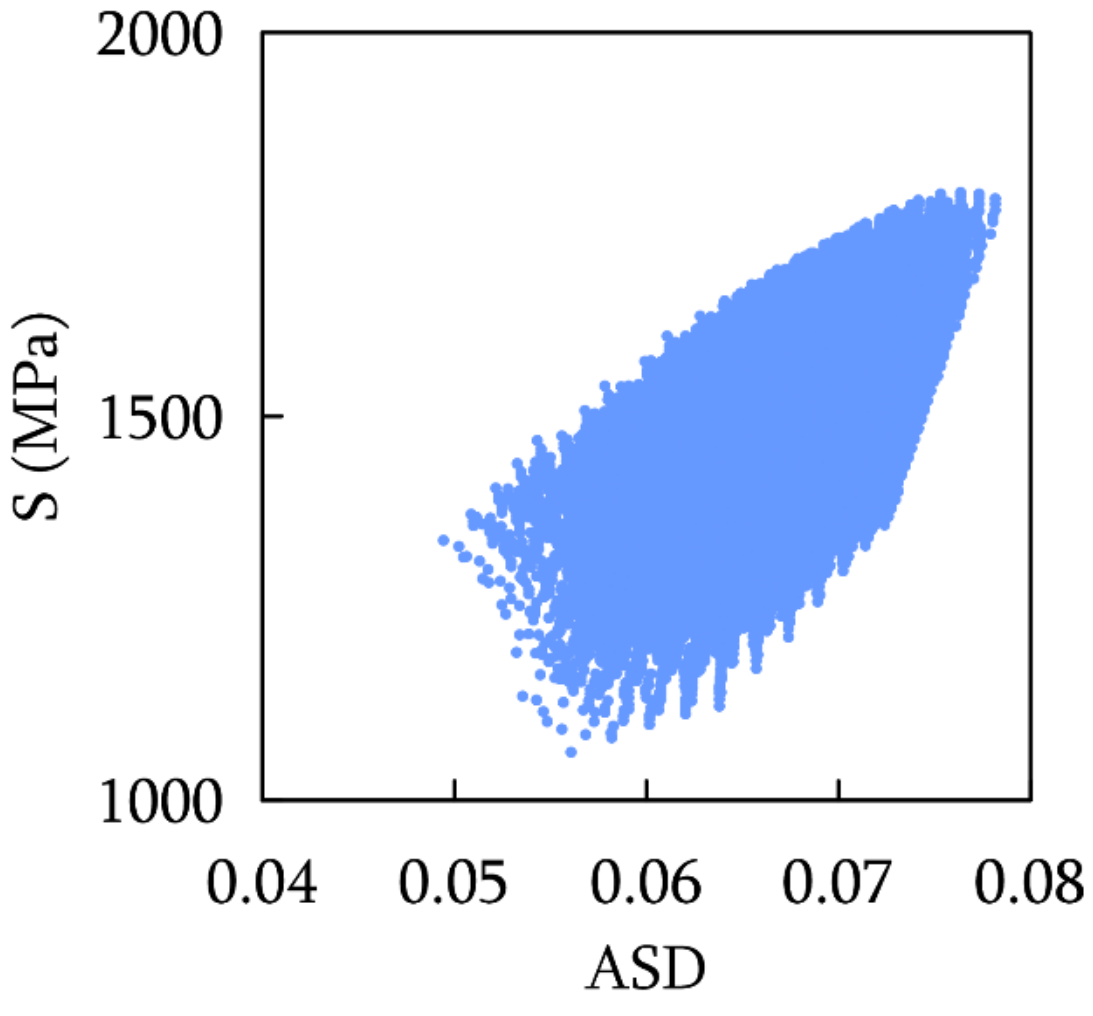

Fig. 7 Relation between ASD and predicted compressive yield stress of designed alloys

According to the results in Figs. 6 and 7, it can be seen that parameters VEC and ASD can affect the strength of RHEAs. Other parameters which can affect the strength of a solid solution alloy are elastic modules [5], lattice distortion [41], electronegativity differences [67] and short range orders [68]. In order to investigate the effect of an alloying element, these parameters should be considered simultaneously. Therefore, the role of a constituent element in changing the strength (Fig. 5) may be explained by considering the role of that element in changing the above parameters and how changing the above parameters can affect the strength (strengthening or softening). It should be mentioned that the results in Fig. 5 are only valid for the alloy system Hf-Mo-Nb-Ta-Ti-V-Zr and within the investigated composition range. With changing the alloy system or the concentration of constituent elements, the effect of constituent elements on strength may change.

The hardest alloy in the Hf-Mo-Nb-Ta-Ti-V-Zr system within the investigated composition range is predicted to be $Hf_{15}Mo_{20}Nb_5Ta_{15}Ti_5V_{20}Zr_{20}$ with the yield stress of 1792 MPa. According to the tensile-ductility data in previous research works [34, 41-52], alloys with tensile yield stresses less than 1200 MPa could have tensile ductility higher than 5% (Fig. 8). Therefore, the developed polynomial is used for finding alloys in the Hf-Mo-Nb-Ta-Ti-V-Zr system with compressive yield stresses less than 1150 MPa.  Around 80 alloys where found and the list of these alloys can be seen in Table 5. Some tensile ductility may be expected for alloys in Table 5. On the other hand, all of the alloys in Table 5 have melting temperatures ($T_m$) higher than 2400 °C (rule of mixture is used for predicting the melting temperatures). Because the loss of high-temperature strength of single-phase bcc RHEs is expected to occur at temperatures higher than 0.6 $T_m$ [69], therefore, it may be predicted that alloys in Table 5 keep their strength up to temperatures of around 1300 °C (0.6 $T_m$) which is high enough for most of the applications. Therefore, reasonable room temperature (ductility) and high temperature mechanical properties may be expected for alloys in Table 5.

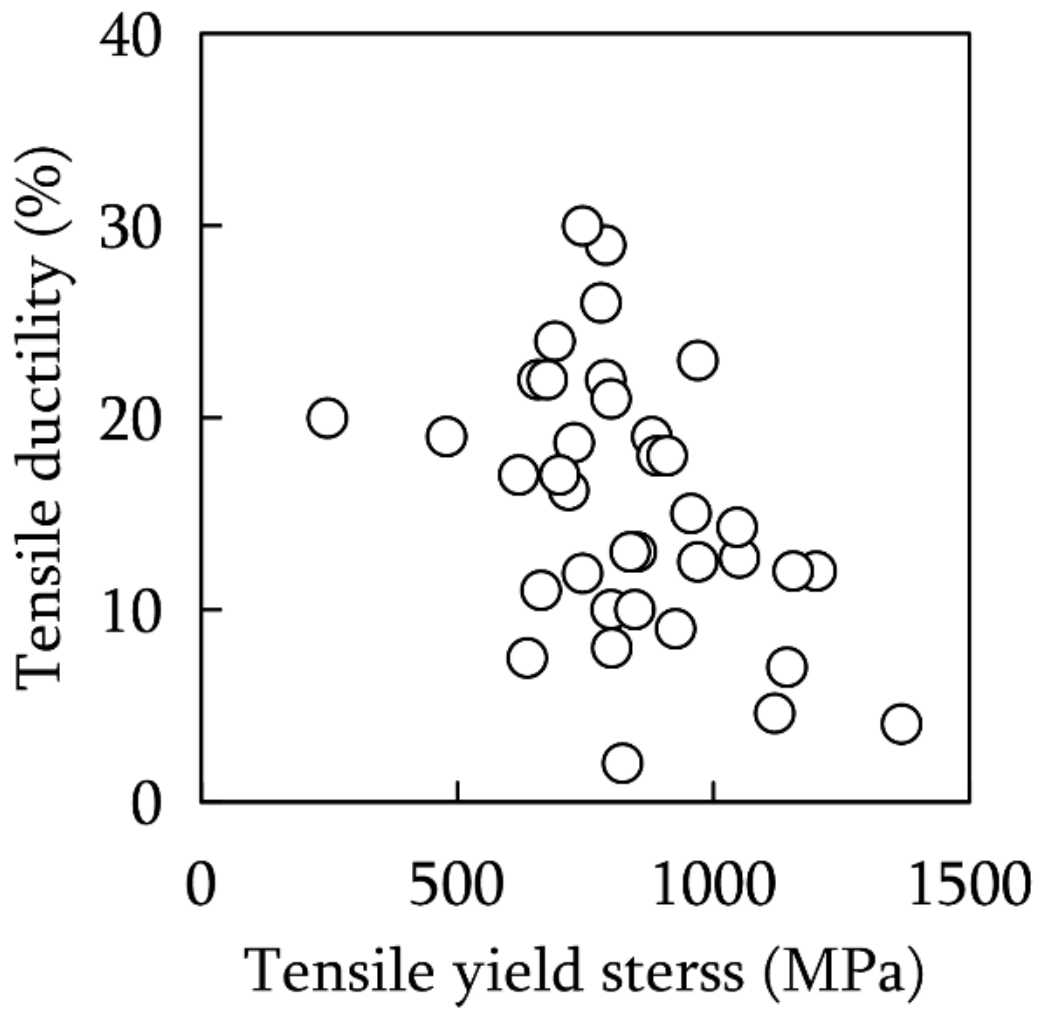

Fig. 8 Tensile ductility versus tensile yield stress for single-phase bcc RHEs [34, 41-52]

Similar to the procedure used for investigating the RHEAs in the Hf-Mo-Nb-Ta-Ti-V-Zr system, other alloy systems (HfMoNbTa, HfNbTiV, etc.) may also be investigated and the strongest alloy in each alloy system can be identified.

Although the predictions may not be very accurate, but they can be considered as good estimates of strength values. Considering the large composition domain which exists within the central regions of multicomponent phase diagrams, and the vast number of alloys which can be selected, even having estimates of strength values can be very valuable. Therefore, the developed polynomial can be considered as a valuable tool in designing of RHEAs.

Table 5. Designed alloys with melting point higher than 2400°C and yield stress less than 1150 MPa (melting points are calculated by the rule of mixture and yield stress values are predicted by developed polynomial). The loss of high-temperature strength of these alloys is predicted to occur at temperatures higher than 0.6 $T_m$ [69]. Tensile ductility higher than 5% is predicted for these alloys (Fig. 8).

| **Alloy** | **Hf** | **Mo** | **Nb** | **Ta** | **Ti** | **V** | **Zr** | **$T_m$ (°C)** | **S(MPa)** |
|---|---|---|---|---|---|---|---|---|---|
| 1 | 10 | 5 | 20 | 20 | 20 | 20 | 5 | 2534.6 | 1120.955 |
| 2 | 12.5 | 5 | 17.5 | 20 | 20 | 20 | 5 | 2528.5 | 1134.21 |
| 3 | 12.5 | 5 | 20 | 17.5 | 20 | 20 | 5 | 2515 | 1128.221 |
| 4 | 12.5 | 5 | 20 | 20 | 17.5 | 20 | 5 | 2548.725 | 1143.108 |
| 5 | 12.5 | 5 | 20 | 20 | 20 | 17.5 | 5 | 2542.675 | 1114.629 |
| 6 | 15 | 5 | 15 | 20 | 20 | 20 | 5 | 2522.4 | 1143.265 |
| 7 | 15 | 5 | 17.5 | 17.5 | 20 | 20 | 5 | 2508.9 | 1137.901 |
| 8 | 15 | 5 | 17.5 | 20 | 20 | 17.5 | 5 | 2536.575 | 1124.309 |
| 9 | 15 | 5 | 20 | 15 | 20 | 20 | 5 | 2495.4 | 1130.925 |
| 10 | 15 | 5 | 20 | 17.5 | 17.5 | 20 | 5 | 2529.125 | 1146.799 |
| 11 | 15 | 5 | 20 | 17.5 | 20 | 17.5 | 5 | 2523.075 | 1118.32 |
| 12 | 15 | 5 | 20 | 20 | 17.5 | 17.5 | 5 | 2556.8 | 1133.206 |
| 13 | 15 | 5 | 20 | 20 | 20 | 12.5 | 7.5 | 2549.375 | 1137.256 |
| 14 | 15 | 5 | 20 | 20 | 20 | 15 | 5 | 2550.75 | 1102.69 |
| 15 | 15 | 7.5 | 20 | 20 | 20 | 12.5 | 5 | 2568.575 | 1146.651 |
| 16 | 17.5 | 5 | 12.5 | 20 | 20 | 20 | 5 | 2516.3 | 1148.12 |
| 17 | 17.5 | 5 | 15 | 17.5 | 20 | 20 | 5 | 2502.8 | 1143.381 |
| 18 | 17.5 | 5 | 15 | 20 | 20 | 17.5 | 5 | 2530.475 | 1129.789 |
| 19 | 17.5 | 5 | 17.5 | 15 | 20 | 20 | 5 | 2489.3 | 1137.03 |
| 20 | 17.5 | 5 | 17.5 | 17.5 | 20 | 17.5 | 5 | 2516.975 | 1124.425 |
| 21 | 17.5 | 5 | 17.5 | 20 | 17.5 | 17.5 | 5 | 2550.7 | 1139.311 |
| 22 | 17.5 | 5 | 17.5 | 20 | 20 | 12.5 | 7.5 | 2543.275 | 1143.361 |
| 23 | 17.5 | 5 | 17.5 | 20 | 20 | 15 | 5 | 2544.65 | 1108.795 |
| 24 | 17.5 | 5 | 20 | 12.5 | 20 | 20 | 5 | 2475.8 | 1129.066 |
| 25 | 17.5 | 5 | 20 | 15 | 17.5 | 20 | 5 | 2509.525 | 1145.928 |
| 26 | 17.5 | 5 | 20 | 15 | 20 | 17.5 | 5 | 2503.475 | 1117.449 |
| 27 | 17.5 | 5 | 20 | 17.5 | 17.5 | 17.5 | 5 | 2537.2 | 1133.323 |
| 28 | 17.5 | 5 | 20 | 17.5 | 20 | 12.5 | 7.5 | 2529.775 | 1137.373 |
| 29 | 17.5 | 5 | 20 | 17.5 | 20 | 15 | 5 | 2531.15 | 1102.806 |
| 30 | 17.5 | 5 | 20 | 20 | 15 | 17.5 | 5 | 2570.925 | 1148.984 |
| 31 | 17.5 | 5 | 20 | 20 | 17.5 | 15 | 5 | 2564.875 | 1117.693 |
| 32 | 17.5 | 5 | 20 | 20 | 20 | 7.5 | 10 | 2556.075 | 1142.234 |
| 33 | 17.5 | 5 | 20 | 20 | 20 | 10 | 7.5 | 2557.45 | 1117.668 |
| 34 | 17.5 | 5 | 20 | 20 | 20 | 12.5 | 5 | 2558.825 | 1085.139 |
| 35 | 17.5 | 7.5 | 20 | 17.5 | 20 | 12.5 | 5 | 2548.975 | 1146.768 |
| 36 | 17.5 | 7.5 | 20 | 20 | 20 | 10 | 5 | 2576.65 | 1127.063 |
| 37 | 20 | 5 | 10 | 20 | 20 | 20 | 5 | 2510.2 | 1148.775 |
| 38 | 20 | 5 | 12.5 | 17.5 | 20 | 20 | 5 | 2496.7 | 1144.661 |
| 39 | 20 | 5 | 12.5 | 20 | 20 | 17.5 | 5 | 2524.375 | 1131.069 |
| 40 | 20 | 5 | 15 | 15 | 20 | 20 | 5 | 2483.2 | 1138.935 |
| 41 | 20 | 5 | 15 | 17.5 | 20 | 17.5 | 5 | 2510.875 | 1126.33 |
| 42 | 20 | 5 | 15 | 20 | 17.5 | 17.5 | 5 | 2544.6 | 1141.216 |
| 43 | 20 | 5 | 15 | 20 | 20 | 12.5 | 7.5 | 2537.175 | 1145.266 |
| 44 | 20 | 5 | 15 | 20 | 20 | 15 | 5 | 2538.55 | 1110.7 |
| 45 | 20 | 5 | 17.5 | 12.5 | 20 | 20 | 5 | 2469.7 | 1131.596 |
| 46 | 20 | 5 | 17.5 | 15 | 17.5 | 20 | 5 | 2503.425 | 1148.458 |
| 47 | 20 | 5 | 17.5 | 15 | 20 | 17.5 | 5 | 2497.375 | 1119.979 |
| 48 | 20 | 5 | 17.5 | 17.5 | 17.5 | 17.5 | 5 | 2531.1 | 1135.853 |
| 49 | 20 | 5 | 17.5 | 17.5 | 20 | 12.5 | 7.5 | 2523.675 | 1139.903 |
| 50 | 20 | 5 | 17.5 | 17.5 | 20 | 15 | 5 | 2525.05 | 1105.336 |
| 51 | 20 | 5 | 17.5 | 20 | 17.5 | 15 | 5 | 2558.775 | 1120.223 |
| 52 | 20 | 5 | 17.5 | 20 | 20 | 7.5 | 10 | 2549.975 | 1144.764 |
| 53 | 20 | 5 | 17.5 | 20 | 20 | 10 | 7.5 | 2551.35 | 1120.198 |
| 54 | 20 | 5 | 17.5 | 20 | 20 | 12.5 | 5 | 2552.725 | 1087.669 |
| 55 | 20 | 5 | 20 | 10 | 20 | 20 | 5 | 2456.2 | 1122.645 |
| 56 | 20 | 5 | 20 | 12.5 | 17.5 | 20 | 5 | 2489.925 | 1140.494 |
| 57 | 20 | 5 | 20 | 12.5 | 20 | 15 | 7.5 | 2482.5 | 1148.619 |
| 58 | 20 | 5 | 20 | 12.5 | 20 | 17.5 | 5 | 2483.875 | 1112.015 |
| 59 | 20 | 5 | 20 | 15 | 17.5 | 17.5 | 5 | 2517.6 | 1128.876 |
| 60 | 20 | 5 | 20 | 15 | 20 | 12.5 | 7.5 | 2510.175 | 1132.926 |
| 61 | 20 | 5 | 20 | 15 | 20 | 15 | 5 | 2511.55 | 1098.36 |
| 62 | 20 | 5 | 20 | 17.5 | 15 | 17.5 | 5 | 2551.325 | 1145.525 |
| 63 | 20 | 5 | 20 | 17.5 | 17.5 | 12.5 | 7.5 | 2543.9 | 1148.8 |
| 64 | 20 | 5 | 20 | 17.5 | 17.5 | 15 | 5 | 2545.275 | 1114.234 |
| 65 | 20 | 5 | 20 | 17.5 | 20 | 7.5 | 10 | 2536.475 | 1138.775 |
| 66 | 20 | 5 | 20 | 17.5 | 20 | 10 | 7.5 | 2537.85 | 1114.209 |
| 67 | 20 | 5 | 20 | 17.5 | 20 | 12.5 | 5 | 2539.225 | 1081.68 |
| 68 | 20 | 5 | 20 | 20 | 15 | 15 | 5 | 2579 | 1129.895 |
| 69 | 20 | 5 | 20 | 20 | 17.5 | 10 | 7.5 | 2571.575 | 1129.095 |
| 70 | 20 | 5 | 20 | 20 | 17.5 | 12.5 | 5 | 2572.95 | 1096.566 |
| 71 | 20 | 5 | 20 | 20 | 20 | 5 | 10 | 2564.15 | 1114.995 |
| 72 | 20 | 5 | 20 | 20 | 20 | 7.5 | 7.5 | 2565.525 | 1092.466 |
| 73 | 20 | 5 | 20 | 20 | 20 | 10 | 5 | 2566.9 | 1061.975 |
| 74 | 20 | 7.5 | 17.5 | 17.5 | 20 | 12.5 | 5 | 2542.875 | 1149.298 |
| 75 | 20 | 7.5 | 17.5 | 20 | 20 | 10 | 5 | 2570.55 | 1129.593 |
| 76 | 20 | 7.5 | 20 | 15 | 20 | 12.5 | 5 | 2529.375 | 1142.321 |
| 77 | 20 | 7.5 | 20 | 17.5 | 20 | 10 | 5 | 2557.05 | 1123.604 |
| 78 | 20 | 7.5 | 20 | 20 | 17.5 | 10 | 5 | 2590.775 | 1138.49 |
| 79 | 20 | 7.5 | 20 | 20 | 20 | 5 | 7.5 | 2583.35 | 1130.315 |
| 80 | 20 | 7.5 | 20 | 20 | 20 | 7.5 | 5 | 2584.725 | 1101.861 |
| 81 | 20 | 10 | 20 | 20 | 20 | 5 | 5 | 2602.55 | 1135.66 |

The proposed approach may also be used for other high entropy alloy systems. For example, the face centered cubic (fcc) solid solution HEAs in the Co-Cr-Fe-Mn-Ni system (Contor alloy system) can be considered. The experimental dataset which is shown in Table 6 can be used in this regard [70]. It was reported that all of the alloys listed in Table 6 have a single phase fcc solid solution after annealing [70] except alloy #1 which contained small amounts (4% volume fraction) of a bcc phase. The proposed polynomial is fitted to the experimental dataset in Table 6 and the obtained coefficients are shown in Table 6. The comparison between the experimental data and predictions is shown in Fig. 9. It can be seen that there is a very good agreement between the experimental and prediction results for hardness which shows that the proposed polynomial can also be used for modeling the strength of HEAs in the Contor system.

Table 6. Experimental data used for the modeling the hardness of alloys in Contor system [70]

<table>
<tr><th rowspan="2">Alloy</th><th colspan="5">Chemical composition (at.%)</th><th rowspan="2">Hardness (GPa)</th></tr>
<tr><th>Co</th><th>Cr</th><th>Fe</th><th>Mn</th><th>Ni</th></tr>
<tr><td>1</td><td>0</td><td>23.8</td><td>24.9</td><td>25.8</td><td>25.5</td><td>2.12</td></tr>
<tr><td>2</td><td>10</td><td>23.1</td><td>22.5</td><td>22.5</td><td>21.9</td><td>2.5</td></tr>
<tr><td>3</td><td>20</td><td>20.3</td><td>20</td><td>20.3</td><td>19.4</td><td>2.52</td></tr>
<tr><td>4</td><td>29.7</td><td>18.1</td><td>17.5</td><td>17.7</td><td>17</td><td>2.53</td></tr>
<tr><td>5</td><td>49.6</td><td>12.8</td><td>12.5</td><td>12.7</td><td>12.4</td><td>2.74</td></tr>
<tr><td>6</td><td>25.1</td><td>0</td><td>25</td><td>25.2</td><td>24.7</td><td>2.07</td></tr>
<tr><td>7</td><td>23.7</td><td>5.2</td><td>23.9</td><td>24.2</td><td>23</td><td>2.41</td></tr>
<tr><td>8</td><td>21.2</td><td>15.3</td><td>21.3</td><td>21.6</td><td>20.6</td><td>2.53</td></tr>
<tr><td>9</td><td>18.7</td><td>25.5</td><td>18.7</td><td>18.9</td><td>18.2</td><td>2.5</td></tr>
<tr><td>10</td><td>24.9</td><td>25.6</td><td>0</td><td>25.1</td><td>24.4</td><td>3.22</td></tr>
<tr><td>11</td><td>22.3</td><td>23</td><td>10</td><td>22.8</td><td>21.9</td><td>2.85</td></tr>
<tr><td>12</td><td>17.4</td><td>18.1</td><td>29.8</td><td>17.7</td><td>17</td><td>2.41</td></tr>
<tr><td>13</td><td>12.7</td><td>13</td><td>49.5</td><td>12.6</td><td>12.2</td><td>1.94</td></tr>
<tr><td>14</td><td>24.9</td><td>25.5</td><td>25.1</td><td>0</td><td>24.5</td><td>2.37</td></tr>
<tr><td>15</td><td>22.4</td><td>23</td><td>22.6</td><td>10.2</td><td>21.8</td><td>2.55</td></tr>
<tr><td>16</td><td>17.4</td><td>17.9</td><td>17.4</td><td>30.3</td><td>17</td><td>2.46</td></tr>
<tr><td>17</td><td>12.6</td><td>12.3</td><td>12.7</td><td>50.2</td><td>12.2</td><td>2.31</td></tr>
<tr><td>18</td><td>10.4</td><td>10.5</td><td>10.4</td><td>9.6</td><td>59.1</td><td>2.94</td></tr>
<tr><td>19</td><td>2.1</td><td>2.1</td><td>2.2</td><td>2.4</td><td>91.2</td><td>1.99</td></tr>
<tr><td colspan="7">Obtained coefficients after fitting the polynomial to experimental data</td></tr>
<tr><td></td><td>Co</td><td>Cr</td><td>Fe</td><td>Mn</td><td colspan="2">Ni</td></tr>
<tr><td>$a_i$</td><td>3.5928</td><td>7.8797</td><td>-2.2453</td><td>2.4492</td><td colspan="2">5.5452</td></tr>
<tr><td>$a_i$</td><td>-1.9858</td><td>-18.1814</td><td>4.1906</td><td>-1.8303</td><td colspan="2">-3.9653</td></tr>
</table>

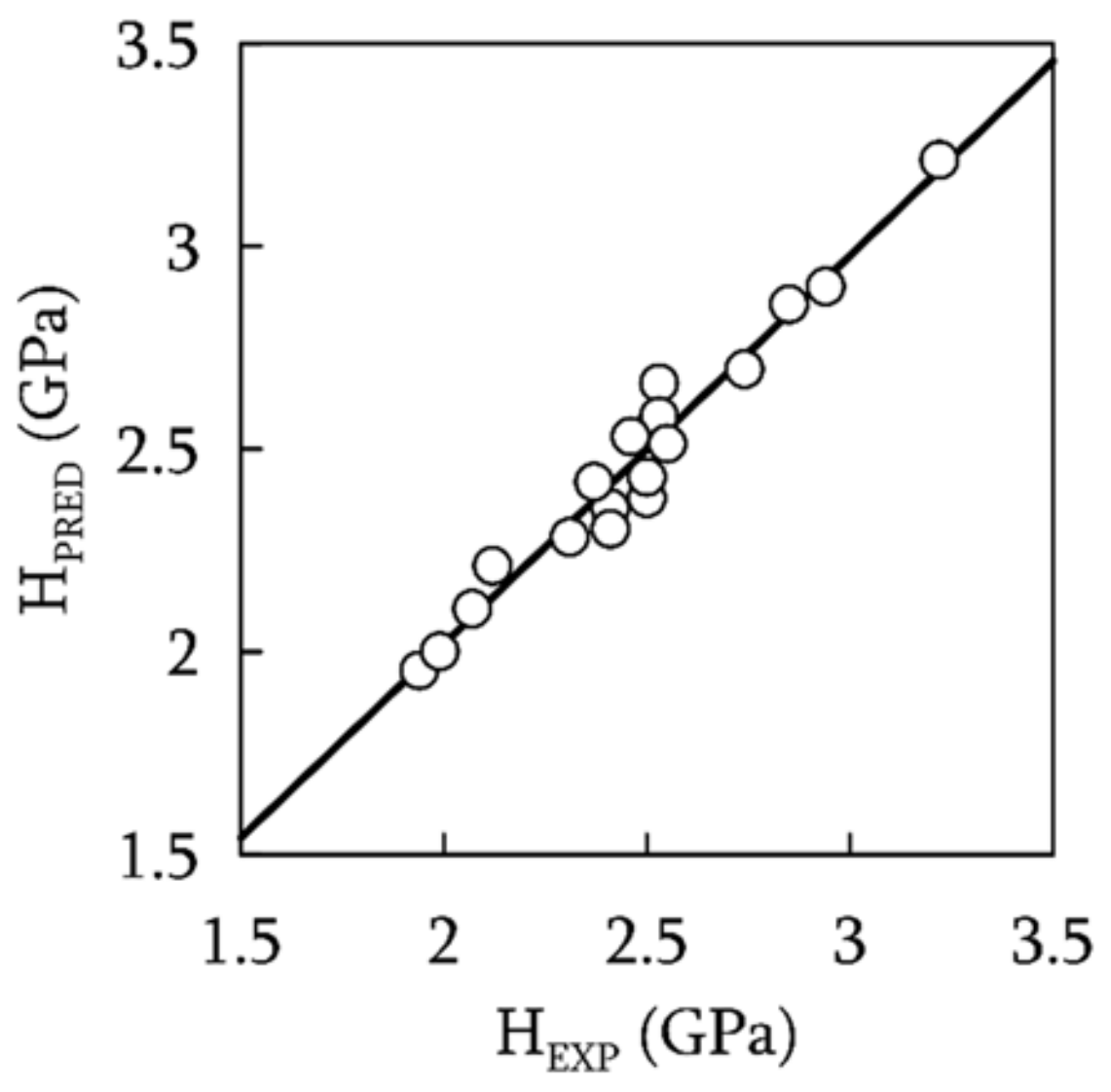

Fig. 9 Comparison between the experimental ($H_{EXP}$) and predictions results ($H_{PRED}$) for the hardness of alloys in Table 6

At the end, few points need to be mentioned about the developed polynomial. (1) The microstructures (including parameters such as the grain size, porosity and micro segregations) of alloys in Table 1 were not the same although all of them were made by the vacuum arc melting technique. Furthermore, the mechanical testing conditions were not exactly the same. So, it may not be accurate to compare the yield stress values in Table 1 just based on the chemical compositions, and microstructural parameters should also be considered. Therefore, it is more reasonable to assume some degree of deviation for the yield stress values in Table 1. If these deviations can be considered, then an accurate polynomial can be obtained. Nevertheless, it was observed that the developed polynomial was able to estimate the yield stress values with a reasonable accuracy by just using the alloy composition as input. If experimental data can be provided for alloys with the same preparation and microstructural conditions, then more accurate results can be obtained. (2) If more experimental data can be provided, then a more accurate polynomial can be developed; furthermore, if more experimental data can be provided then the composition domain (Fig. 1) within which the polynomial is valid can be expanded. According to the relatively fast growing field of RHEAs, more experimental data are expected to be available which could be used for improving the accuracy of the polynomial. Moreover, if reasonable number of data can be provided for alloys containing Cr and W, then the polynomial can be further improved to cover RHEAs alloys containing Cr and W. (3) Other forms of equations may be used for modeling the strength of solid solution alloys, and more accurate results may be obtained. For example, one may develop equations based on parameters $C^{1/2}$ and $C^{3/2}$ (where C shows the concentration) according to the Labusch theory instead of using parameters C and $C^2$.

## 5. Conclusions

In summary, a simple polynomial equation is used to model the compressive yield stress of as-cast solid solution refractory high entropy alloys in the Al-Hf-Mo-Nb-Ta-Ti-V-Zr system. More than 80 experimental data are used for finding the polynomial's coefficients. The results show that the proposed polynomial could model the yield strength of solid solution alloys with a reasonable accuracy. The developed polynomial is used for predicting the strength of RHEAs in the Hf-Mo-Nb-Ta-Ti-V-Zr system. The results show that the strength of designed alloys increases with increasing the value of parameters valence electron concentration (VEC) and atomic size difference (ASD). Furthermore, the results show that for investigated the alloy system, Mo and Zr have both strengthening effect while Ti shows softening effect.